\documentclass[sn-mathphys,pdflatex,twocolumn]{sn-jnl}

\graphicspath{{Figures/}}

\usepackage[numbers]{natbib}

\jyear{2022}%

\theoremstyle{thmstyleone}%
\theoremstyle{thmstyletwo}%

\theoremstyle{thmstylethree}%

\begin{document}

\title[ ]{Universal Early Warning Signals of Phase Transitions in Climate Systems}


\author*[1]{\fnm{Daniel} \sur{Dylewsky}}\email{ddylewsk@uwaterloo.edu}

\author[2]{\fnm{Timothy M.} \sur{Lenton}}\email{T.M.Lenton@exeter.ac.uk}

\author[3]{\fnm{Marten} \sur{Scheffer}}\email{marten.scheffer@wur.nl}

\author[4]{\fnm{Thomas M.} \sur{Bury}}\email{thomas.bury@mcgill.ca}

\author[5]{\fnm{Christopher G.} \sur{Fletcher}}\email{chris.fletcher@uwaterloo.ca}

\author[6]{\fnm{Madhur} \sur{Anand}}\email{manand@uoguelph.ca}

\author[1]{\fnm{Chris T.} \sur{Bauch}}\email{cbauch@uwaterloo.ca}

\affil*[1]{\orgdiv{Department of Applied Mathematics}, \orgname{University of Waterloo}, \orgaddress{\city{Waterloo}, \postcode{N2L 3G1}, \state{ON}, \country{Canada}}}

\affil[2]{\orgdiv{Global Systems Institute}, \orgname{University of Exeter}, \orgaddress{\city{Exeter}, \postcode{EX4 4PY}, \country{United Kingdom}}}

\affil[3]{\orgdiv{Department of Environmental Sciences}, \orgname{Wageningen University}, \orgaddress{\city{Wageningen}, \postcode{6708 PB}, \country{Netherlands}}}

\affil[4]{\orgdiv{Department of Physiology}, \orgname{McGill University}, \orgaddress{\city{Montreal}, \postcode{H3A 0G4}, \state{QC}, \country{Canada}}}

\affil[5]{\orgdiv{Department of Geography and Environmental Management}, \orgname{University of Waterloo}, \orgaddress{\city{Waterloo}, \postcode{N2L 3G1}, \state{ON}, \country{Canada}}}

\affil[6]{\orgdiv{School of Environmental Sciences}, \orgname{University of Guelph}, \orgaddress{\city{Guelph}, \postcode{N1G 2W1}, \state{ON}, \country{Canada}}}


\abstract{The potential for complex systems to exhibit tipping points in which an equilibrium state undergoes a sudden and often irreversible shift is well established, but prediction of these events using standard forecast modeling techniques is quite difficult. This has led to the development of an alternative suite of methods that seek to identify signatures of critical phenomena in data, which are expected to occur in advance of many classes of dynamical bifurcation. Crucially, the manifestations of these critical phenomena are generic across a variety of systems, meaning that data-intensive deep learning methods can be trained on (abundant) synthetic data and plausibly prove effective when transferred to (more limited) empirical data sets. This paper provides a proof of concept for this approach as applied to lattice phase transitions: a deep neural network trained exclusively on 2D Ising model phase transitions is tested on a number of real and simulated climate systems with considerable success. Its accuracy frequently surpasses that of conventional statistical indicators, with performance shown to be consistently improved by the inclusion of spatial indicators. Tools such as this may offer valuable insight into climate tipping events, as remote sensing measurements provide increasingly abundant data on complex geospatially-resolved Earth systems.}


\maketitle

Earth's climate, a vast and rich dynamical system of extraordinary complexity, is experiencing an abrupt and strengthening perturbation due to the effects of anthropogenic climate change. Consequently, climate scientists are faced with the task of extrapolating forecasts into an as yet unobserved parameter regime. Traditional modeling approaches offer valuable insight into possible climate trajectories over the coming decades and centuries, but since they can only be empirically validated on currently-available data, their out-of-sample predictions for a warming world must be treated with caution. This shortcoming is particularly evident with regard to the characterization of tipping points. Many climate models suggest the possibility of tipping phenomena in which some gradual forcing on a system induces a strong feedback effect, resulting in an abrupt and often irreversible shift in the equilibrium state \cite{Lenton2008,Barnosky2012}. The specific conditions under which such an event will occur are difficult to determine, however; the prediction of any given model cannot be concretely validated until it has already come to pass. Understanding the landscape of tipping thresholds we are poised to cross as the Earth warms is crucial for mitigation efforts \cite{Martin2015}. The sudden and irreversible nature of these events means that they should be of the highest priority of any undertaking to reduce harm to ecosystems and human populations, but this requires a more sophisticated advanced warning system than standard climate models can provide.

Recent work to fulfill this need has led to an alternate forecasting framework which aims to detect ``early-warning signals" (EWS) which present in advance of the onset of tipping points. The theoretical basis for this approach lies in bifurcation theory, which suggests that tipping events across a wide variety of systems are characterized by universal behavioral signatures. Most notable among these is the phenomenon of ``critical slowing down" in equilibrium dynamics \cite{Wissel1984,Scheffer2009}, where a system's recovery from small perturbations slows as a bifurcation point is approached, but many others have been proposed \cite{Guttal2008,Dakos2013,Williamson2015}. Statistical indicators commonly used to detect critical slowing in data include spectral methods \cite{Livina2007}, lag-1 autocorrelation \cite{Held2004,Dakos2008}, and variance \cite{Carpenter2006}.

In aggregate, these methods have had mixed success in detecting oncoming abrupt shifts in empirical data from ecological and climate systems. In many cases one or more EWS indicators have been shown to exhibit significant anomaly leading up to transition events in historical records \cite{Alibakhshi2017,Boers2018,Hennekam2020} as well as in controlled lab and field experiments \cite{Drake2010,Carpenter2011,Dai2012,Kramer2012,Dai2013}. However, a number of possible confounding mechanisms  have been identified which might preclude the detection of early-warning signals in a given data set. An abrupt change in some measured variable may not correspond to a bifurcation in the underlying dynamics, and even if it does, the measured signal is not guaranteed to carry hallmarks of critical slowing down \cite{VanderBolt2021,Boerlijst2013,Negahbani2016}. Methods for EWS detection are further complicated by the relative rarity of bifurcation transitions in systems of interest; statistical models are typically calibrated on data sets selected because they are known to be prone to such events, leading to a ``prosecutor's fallacy" bias which can produce an inflated rate of false-positive classifications \cite{Boettiger2012}.

One major asset of EWS detection methods is their universality: whereas traditional modeling techniques struggle to reliably make out-of-sample predictions, approaches based on indicators of critical phenomena are predicted to be transferrable to any comparable instance of a tipping point, even in a system or parameter regime not previously observed. This property circumvents the challenge of time series forecasting: there is no need to construct a future trajectory to predict the likelihood of an oncoming irreversible shift if it can instead be inferred from present-day variability. Moreover, the ability to apply a generic set of criteria for EWS detection to any system offers a great deal of flexibility in model development. Modern machine learning methods for statistical inference often rely on substantial quantities of training data, which can present an obstacle for applications to systems where empirical measurements are not sufficiently abundant. However, if a simulation model can be designed to replicate the relevant properties of the system in question, a training set can be generated to arbitrary proportion. In the case of tipping phenomena, any system which exhibits the predicted critical phenomena can be used to train a model applicable to a wide variety of test cases. Recent results in deep learning methods for EWS detection have supported this claim, demonstrating that models trained on simple synthetic dynamics can be successfully transferred to empirical climate data from various sources \cite{Bury2021,Deb2021}. 

While bifurcations can occur on low-dimensional manifolds in climate systems, this is not the only mechanism by which abrupt transitions might take place. Higher-dimensional, spatially organized systems admit discontinuous shifts of equilibrium which are described by the theoretical framework of phase transitions. These phenomena bear some relationship to dynamical bifurcations (in some cases mean-field approximations provide a reduced form for a system undergoing a phase transition which is equivalent to the normal form for a known bifurcation), but can be more varied in their manifestation and produce different early-warning signals leading up to a transition \cite{Bose2019,Hagstrom2021}. This added diversity presents an added challenge for the task of EWS detection, but shows promise as a resource for mitigating many of the aforementioned shortcomings of bifurcation-based approaches and producing more robust and generalizable classifier models. Recent theoretical work supports the hypothesis that early warning-signal detection in the phase transition framework may present the opportunity to leverage more modern, data-intensive computational techniques to improve performance \cite{Hagstrom2021}. Spatially organized dynamical systems undergoing phase transitions exhibit the same critical slowing down observed in ODE systems approaching local bifurcation, but can also display a variety of spatially resolved critical phenomena including increased range of spatial correlations, front propagation, and emergence of different signatures in the spatial Fourier spectrum \cite{Kefi2014,Bel2012,Rietkerk2021}. Evidence of these spatial indicators has been observed in studies of abrupt transition events observed in remotely sensed ecosystem data \cite{Eby2017,Majumder2019}.

This paper presents a practical proof of concept for the detection of early-warning signals of an oncoming phase transition using indicators for critical phenomena which are expected to generalize across a wide class of systems. Specifically, we demonstrate a deep learning model trained on phase transitions simulated on a 2D square Ising lattice which successfully identifies upcoming phase transitions even when applied to data from other, out-of-sample sources. Deep neural networks, whose capacities for time-series classification are now well-established \cite{Fawaz2019}, have recently been shown to excel in bifurcation theory-based approaches to early warning signal detection \cite{Bury2021,Deb2021}. We base the work of this paper on the hypothesis that these successes will be reproduced and perhaps amplified in a spatiotemporal phase transition framework. Neural networks have great aptitude for information-preserving dimensionality reduction (e.g. in autoencoders) and for detection of subtle patterns in multivariate, spatially organized data streams (e.g. in convolutional networks). The model we present leverages these capabilities to extract more information from known EWS indicators such as variance and autocorrelation than traditional methods could. Coupled with the demonstrated success of deep learning on early-warning signal detection for scalar time series, the shift to models for spatiotemporal data seems quite natural. While much work remains to fully ascertain the potential of these models, the results presented in this paper offer unambiguous confirmation of deep neural early-warning signal detection for dynamical phase transitions.

\section*{Results}

\subsection*{Classification of withheld Ising test data}

\begin{figure*}[!ht]
\centering
\includegraphics[width=\linewidth]{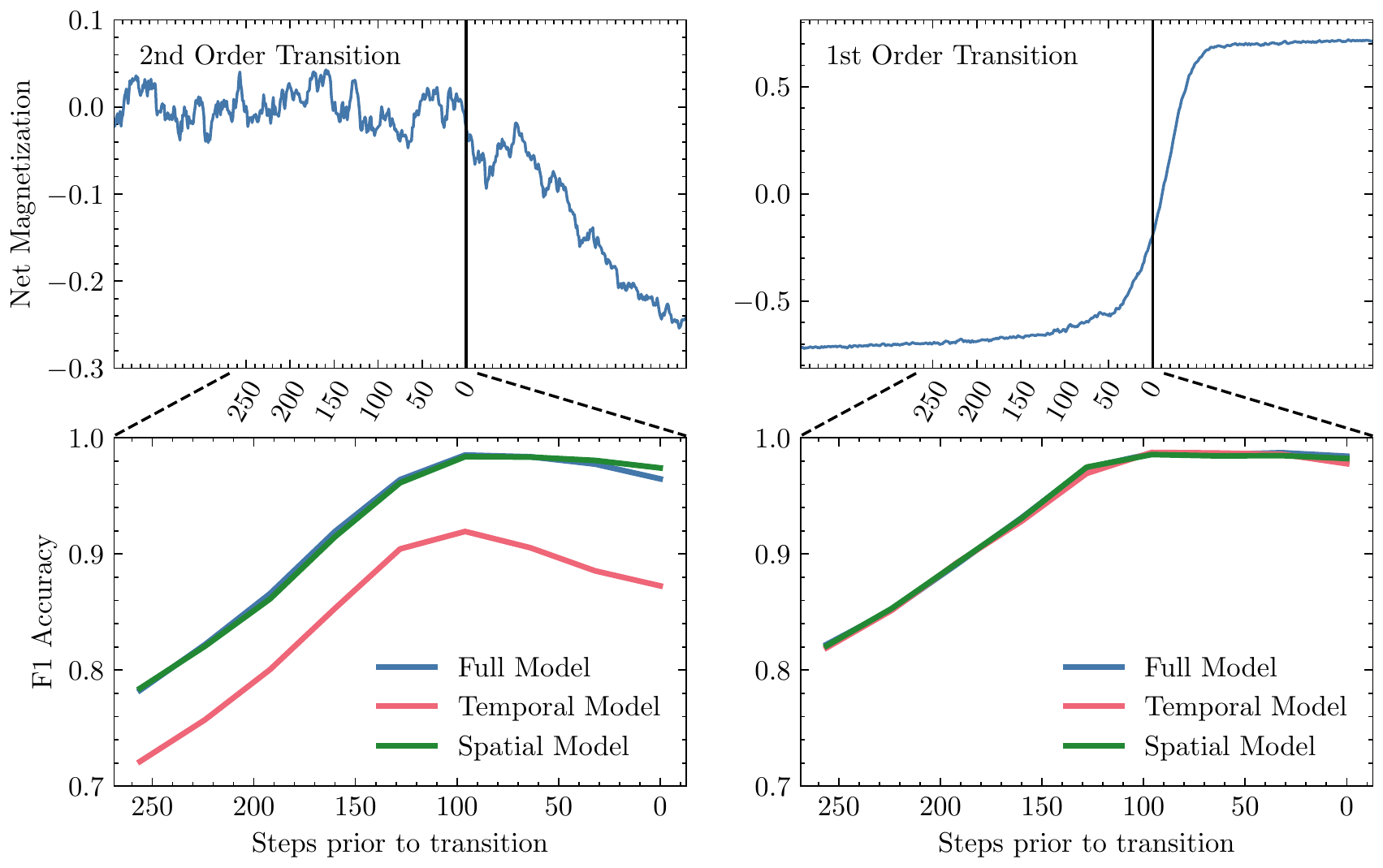}
\caption{Top: Net magnetization of a sample run of an Ising phase transition. Bottom: Classification accuracy of the CNN-LSTM models (computed as F1-score), applied to withheld Ising test data at varying lengths of time prior to the transition. Results are presented for second-order (left) and first-order (right) phase transitions. The vertical black line denotes the identified transition time (computed analytically for second-order, inferred from data for first-order). Results are presented for models trained on data containing all EWS indicators ("Full"), as well as ones trained exclusively on the spatial or temporal indicators.}
\label{fig:var_ls_Ising_combine}
\end{figure*}

Trained models were first tested on Ising model data which had been withheld from the training set. They proved extremely effective in classifying this data: the model trained only on first-order Ising phase transitions achieved a classification accuracy of 96\%, and the one trained on second-order transitions reached above 99\%. Models trained only on spatial or temporal statistical indicators (rather than both together) also performed well, with first- and second-order spatial models at 93\% and 99\%, respectively, and corresponding temporal models at 94\% and 95\%, respectively. Of greater interest for the purposes of early-warning signal detection is the models' forecasting horizon: they must be capable of identifying signatures of critical phenomena \textit{before} the critical point is reached. To this end, models were tested on time series truncated from 0 to 250 time steps prior to the phase transition and evaluated by their F1-score accuracy, which synthesizes measures of precision and recall on classification results from the test set. As shown in Fig. \ref{fig:var_ls_Ising_combine}, the F1-score for the full model on both first- and second-order transitions remains greater than 0.95 out to nearly 150 steps lead time. We note that this model was trained on data truncated 100 steps before the transition, which accounts for the counterintuitive dip in accuracy for low truncation values in the second-order case. This choice for the training set was based on the observation that beyond 100, further truncations stopped resulting in significant improvements of predictive ability.

The units of time and distance in these simulations are arbitrary, as are those in the other phase transition models we use as test cases. In all cases, the presence of useful information for early warning signal detection requires only that the domain of observation be large compared to the characteristic scales of the system's dynamics, which in turn must be large compared to the sampling resolution. These requirements also apply when the model is transferred to real-world data, though it is otherwise agnostic of units of measurement. In training, the 100-step lead time is important to increase the model's sensitivity: critical phenomena are expected to ramp up in intensity as the system approaches a transition, so the model is forced to classify runs based on comparatively weak signatures (and generally performs with minimal depreciation of accuracy when exposed to data from closer to the transition). Note that this truncation is only applied during the training phase---all tests presented in the remainder of this paper are conducted by applying this trained model to time series truncated immediately before the observed abrupt transition.

\subsection*{Comparison with traditional EWS}

Historically, early warning signal detection for tipping events has typically relied on statistical indicators for critical slowing. Tipping events associated with local bifurcations (as well as second- and likely also first-order phase transitions \cite{Hagstrom2021}) are expected to evince longer recovery times from perturbations as a critical threshold is approached, which can be measured with statistics such as variance and lag-1 autocorrelation. Spatiotemporal systems, similarly, often exhibit an increase in spatial correlation length as they near a tipping point. Conventional methods for EWS detection look for anomalies in these statistics as evidence of an oncoming tipping event. To compare our neural model to these traditional approaches, we apply both approaches to the same Ising model data set.

Classification for the traditional indicators was carried out by applying a discrimination threshold to the Kendall $\tau$ statistic, used as a robust proxy for the strength of monotonic trends, following the example of \cite{Bury2021}. In almost all cases the neural model offers improved classification accuracy over the other tested statistics. This result is consistent with those of other deep learning models \cite{Bury2021,Deb2021}, and is hypothesized to be owed to neural networks' capacity for more subtle pattern recognition and synthesis of simultaneous evidence from multiple indicators.

\begin{figure}[!ht]
\centering
\includegraphics[width=\columnwidth]{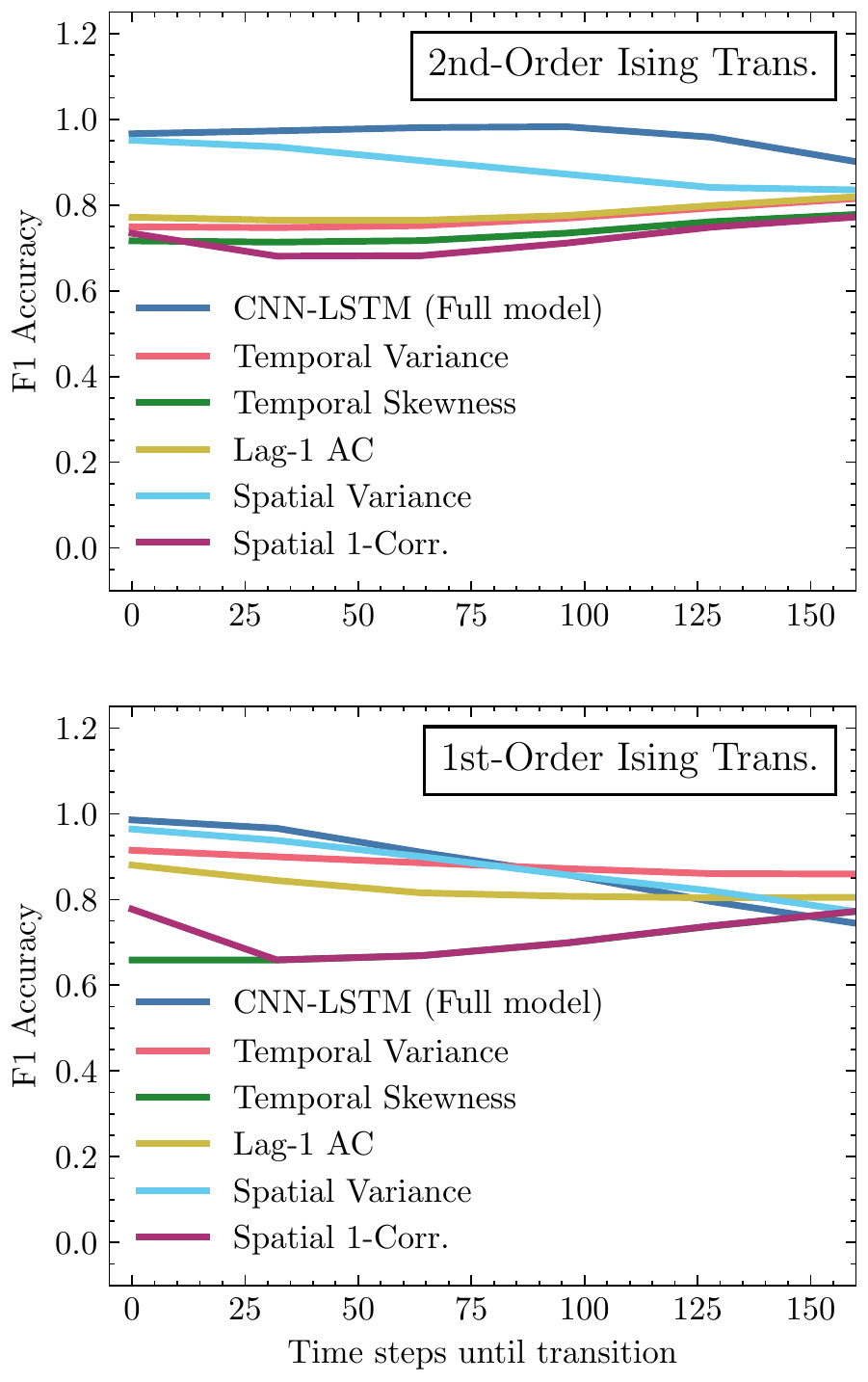}
\caption{Classification accuracy (computed as F1-score) for the CNN-LSTM model and for traditional EWS statistics (using Kendall $\tau$ discrimination threshold) applied to second-order (left) and first-order (right) Ising model simulations at varied time intervals prior to transition. For readability only 5 of the 12 computed traditional indicators are plotted, but in all cases the accuracy of those omitted is comparable to (or lower than) that of those included.}
\label{fig:var_ls_Ising_combined_classical}
\end{figure}

The classification accuracies of the neural models trained on first- and second-order Ising transitions are compared to those of five traditional EWS indicators in Fig. \ref{fig:var_ls_Ising_combined_classical}, tested on runs withheld from their respective training sets. Here, the CNN-LSTM classifiers outperform their traditional counterparts at all observed lead times to transition. We grant that this is perhaps not an entirely fair comparison: the neural models have access to all of these indicators simultaneously, whereas the traditional classifiers can only use one at a time. Indeed, it has been shown that a composite methodology incorporating more than one of these statistics can improve early warning forecasts \cite{Clements2019}. We have refrained from a more in-depth analysis for the time being, however, as the aim of this paper is to present a proof of concept for a new methodology for detection of tipping points in spatiotemporal systems. Although the performance statistics in Fig. \ref{fig:var_ls_Ising_combined_classical} are encouraging, considering that the Ising-trained CNN-LSTM represents a new and unrefined approach to the problem, a proper comparison would require considerable fine-tuning of both techniques, which we leave to future work.

\subsection*{Classification of out-of-sample test models}

Results are next presented for tests on other phase transition models. The models' success on withheld Ising data does not guarantee that they have learned to identify the expected critical phenomena which are generalizable to other tipping events. To evaluate this, data sets are prepared for alternate models known to undergo phase transitions. We have selected two sample models, with the criteria that they evolve on a two-dimensional lattice and hold relevance to climate applications. The first is a coupled vegetation-water model, adapted from Dakos et. al. (2011) \cite{Dakos2011}, in which local positive feedback dynamics facilitate a desertification transition that sees vegetation density drop abruptly when rainfall falls below a critical threshold. The second is a sea ice percolation model, inspired by work demonstrating how brine transport in porous sea ice can undergo a phase transition known to occur in percolation models \cite{Golden1998}. Details on these models are provided in Appendix A.

\begin{figure*}[!ht]
\centering
\includegraphics[width=\linewidth]{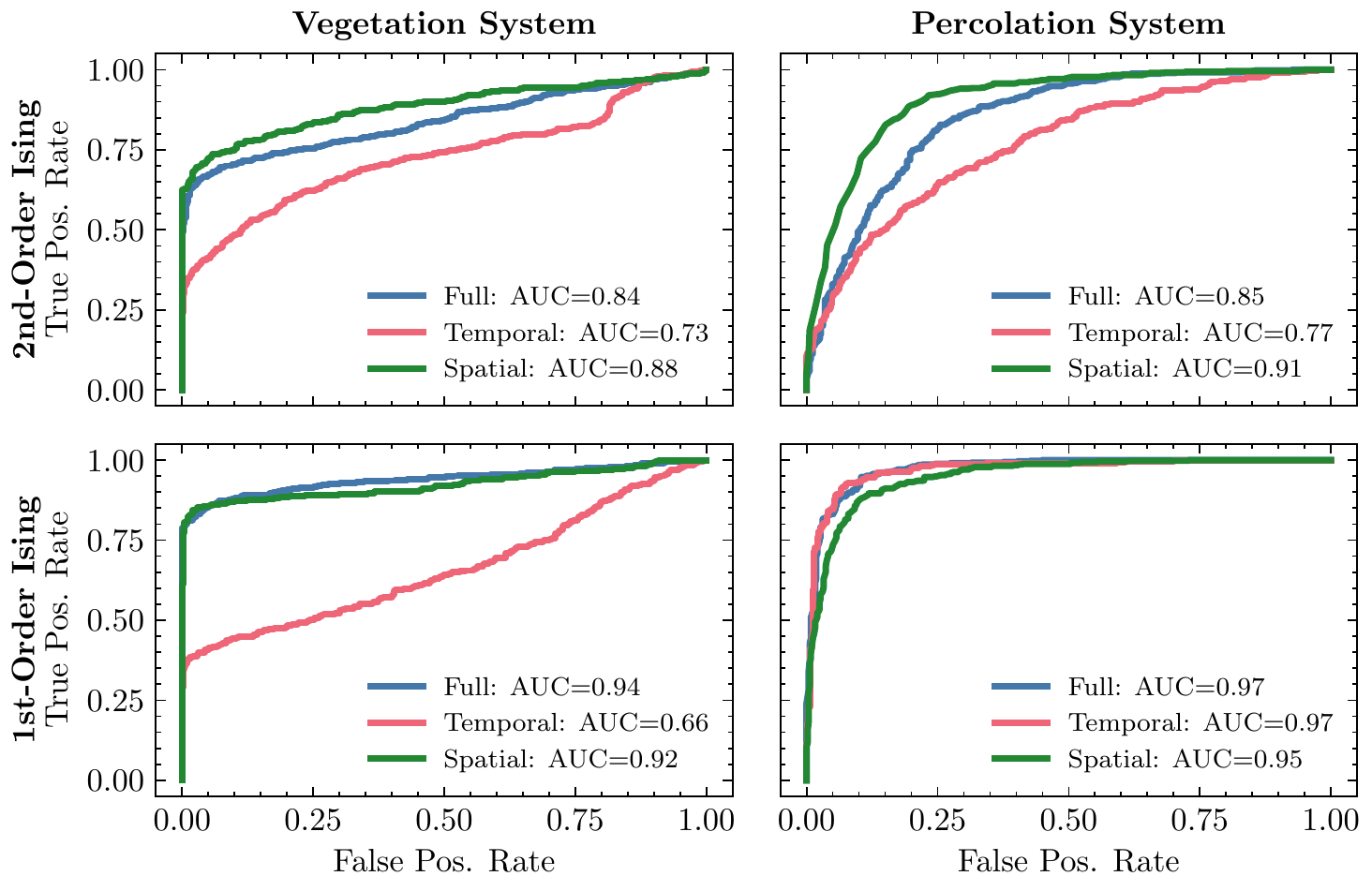}
\caption{Classification results for the neural models trained on second-order Ising transitions (top) and first-order Ising transitions (bottom) applied to out-of-sample test data. Left: coupled vegetation-water model, right: sea ice percolation model.}
\label{fig:Kefi_perc_ROC_combined_HP_1}
\end{figure*}

Data sets are prepared similarly to the Ising training data: an externally-controlled parameter is varied linearly, either passing through its critical value (transition runs) or held far away from it (null runs). The CNN-LSTM models trained purely on Ising transitions are tested on this data to evaluate their capacity for generalization. Receiver operating characteristic (ROC) curves are generated by applying a variable discrimination threshold directly to the values from the classification output layer of the neural model. Results, presented in Fig. \ref{fig:Kefi_perc_ROC_combined_HP_1}, offer evidence that the models have successfully learned generic heuristics to detect oncoming transitions: all tests yielded an area under ROC curve well above baseline (AUC = 0.5 for randomized classification), with peak values of 0.94 and 0.97 for the vegetation and percolation systems, respectively.

\subsection*{Classification of abrupt shifts in climate systems}
\begin{figure}[!ht]
\centering
\includegraphics[width=\columnwidth]{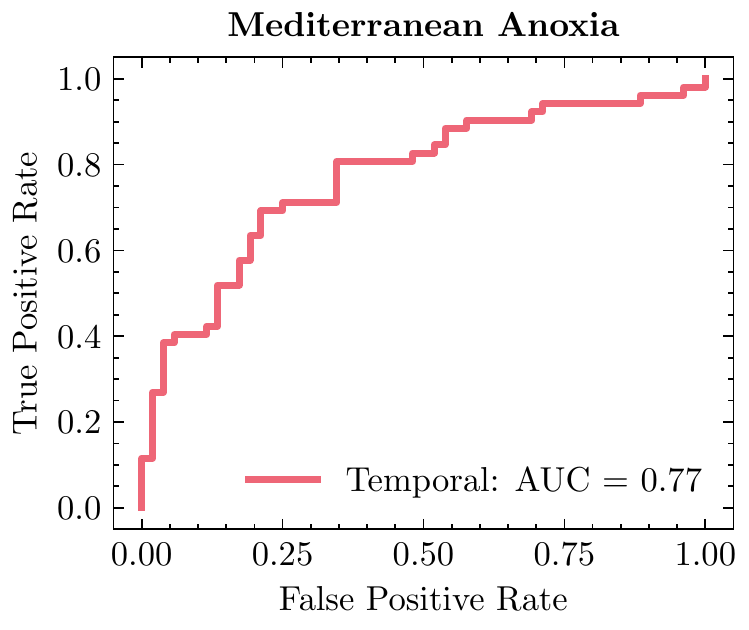}
\caption{Classification results for empirical data on Mediterranean anoxia events using a model trained on second-order Ising transitions. Only temporal results are available because the source data did not have any spatial component.}
\label{fig:anoxia_ROC_curve}
\end{figure}

The final test cases we present are those making use of empirical climate data. Fig. \ref{fig:anoxia_ROC_curve} shows results for a data set drawn from sedimentary records of anoxia events in the eastern Mediterranean which has previously been the subject of EWS analysis using traditional statistical methods \cite{Hennekam2020}. Observed transitions are compared with null runs generated using AR(1) processes defined using statistics drawn from pre-transition dynamics in the measured data. This example illustrates the generalization of our model, trained for recognition of spatiotemporal tipping events, to purely temporal time series. Even lacking access to the spatial features which often considerably improve its performance, the neural model achieves passable success in distinguishing upcoming transition events (though it does not perform as well as a similar deep learning model designed for scalar time series and tested on the same data \cite{Bury2021}). 

A second set of practical test cases are drawn from the simulations aggregated by the Coupled Model Intercomparison Project 5 (CMIP5), which provide spatio-temporal gridded outputs for a full range of state and flux variables from the major components of the Earth system. A catalog of abrupt transitions published by Bathiany et. al. \cite{Bathiany2020} is used to construct a data set of test material for tipping point detection. Analysis is restricted to variables for which two-dimensional geospatial data is available, sampled at a monthly temporal resolution and simulated in the RCP8.5 scenario of future greenhouse gas emissions. Each time series is downsampled by month to omit annual oscillation phenomena, resulting in a data set containing 12 times as many time series, sampled yearly. Transition (and null) runs are designated by high (low) abruptness scores as computed by the Bathiany edge detection method. These scores are computed for each point on the spatial lattice, so we are able to isolate spatiotemporal measurements associated with the immediate geographic vicinity of any flagged transition. Of the many models included in CMIP5, we have selected four test candidates based on their high incidence of abrupt shifts as measured by the Bathiany metric: GFDL-ESM2M, FGOALS-g2, CSIRO-Mk3-6-0, and CMCC-CESM. This is admittedly a somewhat artificial selection criterion in the sense that the number of tipping events in the test data is inflated relative to what would be expected in generic real data, but for evaluation purposes we prioritize having sufficiently large class populations to gather useful statistics. Practical deployment of the model on real-world data would likely require some additional tuning to suppress the rate of false positive classifications, although it is our hope that with continued development our deep learning approach capable of synthesizing multiple indicators will prove more naturally robust to these errors than traditional EWS methods.

\begin{figure*}[!ht]
\centering
\includegraphics[width=\linewidth]{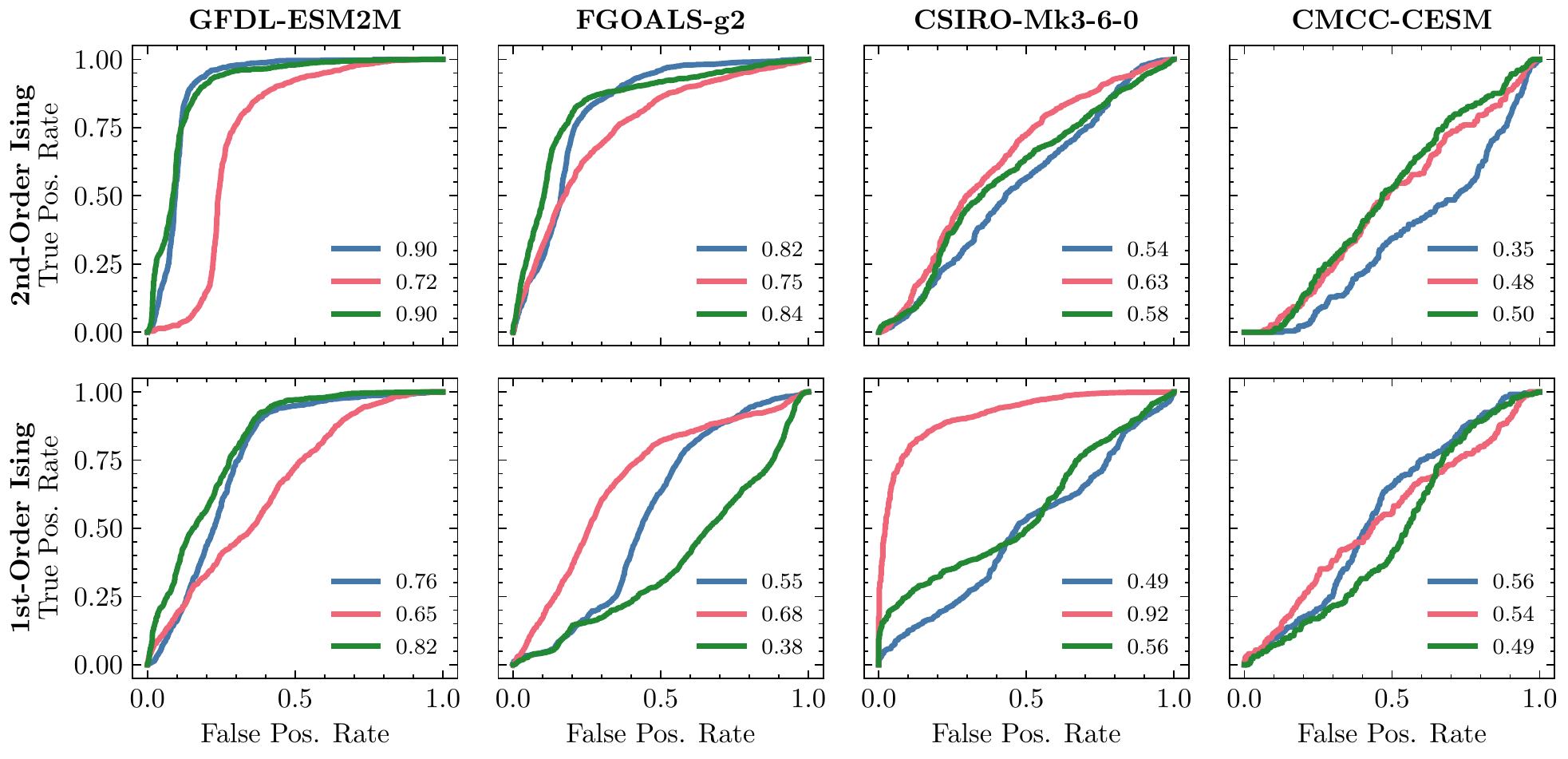}
\caption{ROC curves for classification of abrupt transitions in four models from the CMIP5 project using CNN-LSTM models trained on second-order (top) and first-order (bottom) Ising transitions. Results are presented for the full, temporal, and spatial classifiers in blue, pink, and green, respectively. Areas under each curve are annotated in the legends.}
\label{fig:CMIP_ROC_combined}
\end{figure*}

\begin{figure}[!ht]
\centering
\includegraphics[width=\linewidth]{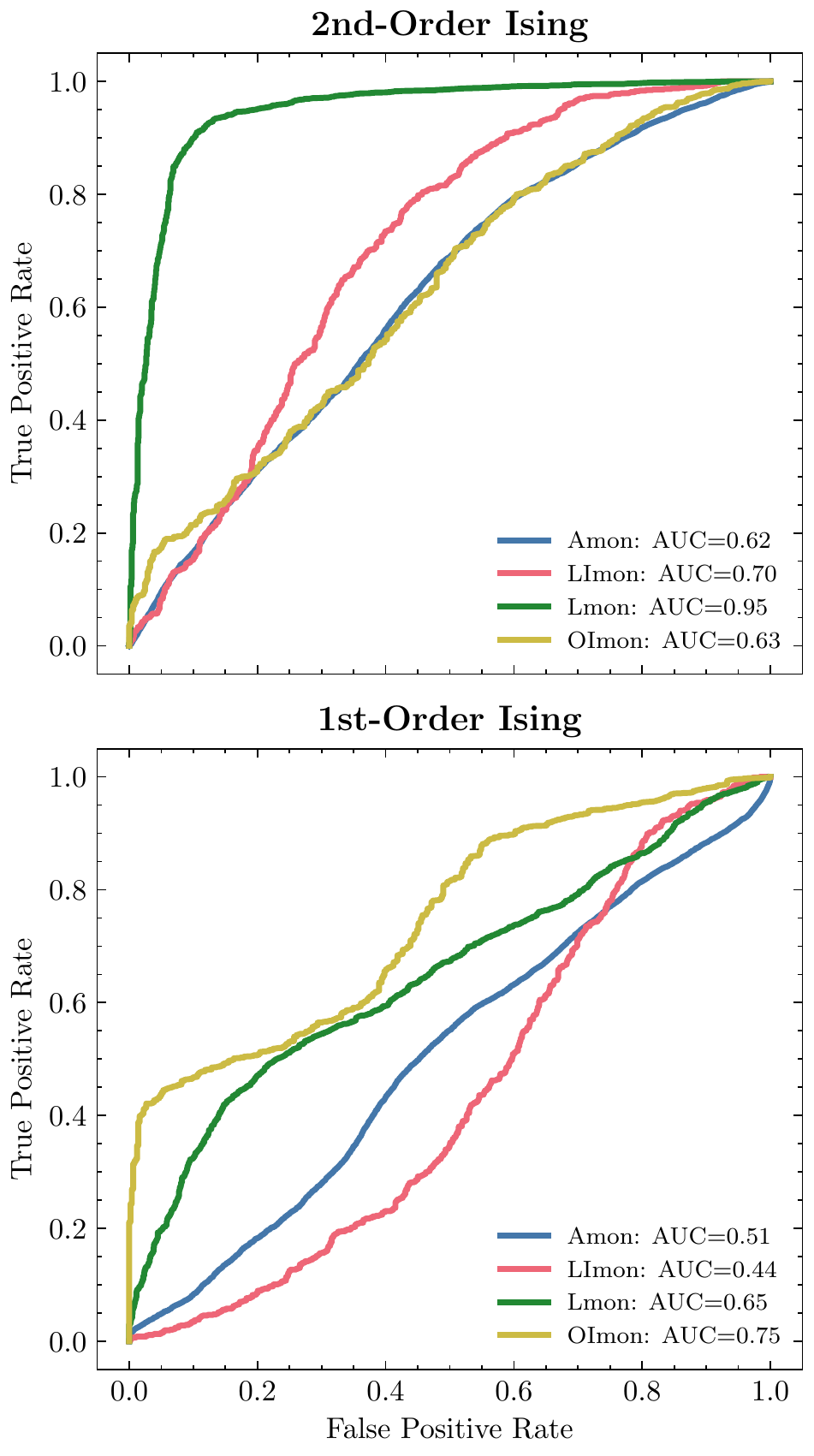}
\caption{ROC curves for classification (using all spatial and temporal coordinates) of CMIP5 data (four global climate models, aggregated) with variables separated by climate subsystem: Amon (Monthly Mean Atmospheric Fields and Some Surface Fields), LImon (Monthly Mean Land Cryosphere Fields), Lmon (Monthly Mean Land Fields, Including Physical, Vegetation, Soil, and Biogeochemical Variables), and OImon (Monthly Mean Ocean Cryosphere Fields)}
\label{fig:CMIP_ROC_table}
\end{figure}

Classification results for this data set, separated by model, are presented in Fig. \ref{fig:CMIP_ROC_combined}. The outcomes of these tests are mixed; with area under the ROC curve ranging from 90\% to below 50\%. This is not unexpected, given the methodology: the positive classification group was constructed from abrupt shifts observed in the data, without any further insight into the underlying dynamics. Complex, multiscale models such as those used in these sophisticated climate simulations often admit tipping points born of induced dynamical feedback effects, but they are also entirely capable of producing sudden shifts without any underpinning critical nature (e.g. when rare extreme events arise from intrinsic dynamics). It is therefore quite plausible that many of the included transition events are not preceded by any signatures of critical phenomena, placing them outside the scope of EWS detection methods. The CMIP5 models analyzed vary in their domains, mechanisms, and intrinsic time and length scales, so it stands to reason that some should be more prone to instances of true critical transitions than others (as is empirically suggested by the results of Bathiany et. al.). 

\subsection*{Critical phenomena of phase transitions}
Hagstrom and Levin (2021) \cite{Hagstrom2021} offer a theoretical discussion of the predicted manifestation of early warning signals in first- and second-order phase transitions. Of central concern to this analysis is the observation that bifurcation analysis of the free energy landscape suggests that critical slowing should take place for second-order transitions, but not first-order. In practice, however, first-order transitions are observed to reliably exhibit the same EWS behaviors. Understanding this contradiction is essential to the application of our results, as it is first-order phase transitions which correspond to the tipping points of greatest concern in climate systems: these are the phase transitions which result in a sudden, discontinuous change in equilibrium state, exhibit hysteresis, and cannot be easily reversed. Hagstrom and Levin propose a resolution to this tension using the concept of spinodal instabilities, which occur when an out-of-equlibrium system in a metastable state loses its local stability. They argue that this phenomenon, which often occurs along the trajectory of a system traversing from its old (subcritical) global minimum of free energy to a new (supercritical) global minimum, results in the presence of early-warning indicators similar to those which characterize second-order transitions.

Spinodal instability occurs when a homogeneous phase becomes thermodynamically unstable, resulting in a spontaneous, spatially distributed separation \cite{Binder1987}. This phenomenon is contrasted with the more common mechanism in which phase separation is initiated by localized nucleation events resulting from thermal fluctuations. Spinodal criticality is rarely observed in real physical systems, which at nonzero temperature are subject to transition by the latter mechanism before they can reach true spinodal instability \cite{Hagstrom2021}. The same constraint applies to the simulations carried out in this paper, both of the first-order Ising model (at constant, finite temperature) and of the other phase transition systems---transitions are observed to take place primarily by means of nucleation. This does not mean, however, that early warning signals associated with first-order phase transitions are necessarily beyond the reach of the neural model. EWS detection methods rely on the core principle that critical phenomena can be observed well before a system actually crosses the threshold of a critical transition, and this holds even if that transition is approached but never reached. Power-law behavior associated with spinodal criticality has been demonstrated in fiber bundle models well before the system reaches its spinodal point \cite{Hagstrom2021}, and the same ought to apply for thermal phase transitions (at least within some low temperature regime). The performance of the neural model trained on first-order Ising data serves to test the applicability of this conclusion, offering evidence for a variety of test systems of whether hallmarks of spinodal criticality can be detected in circumstances (somewhat) resembling real-world conditions.

\begin{figure}[!ht]
\centering
\includegraphics[width=\columnwidth]{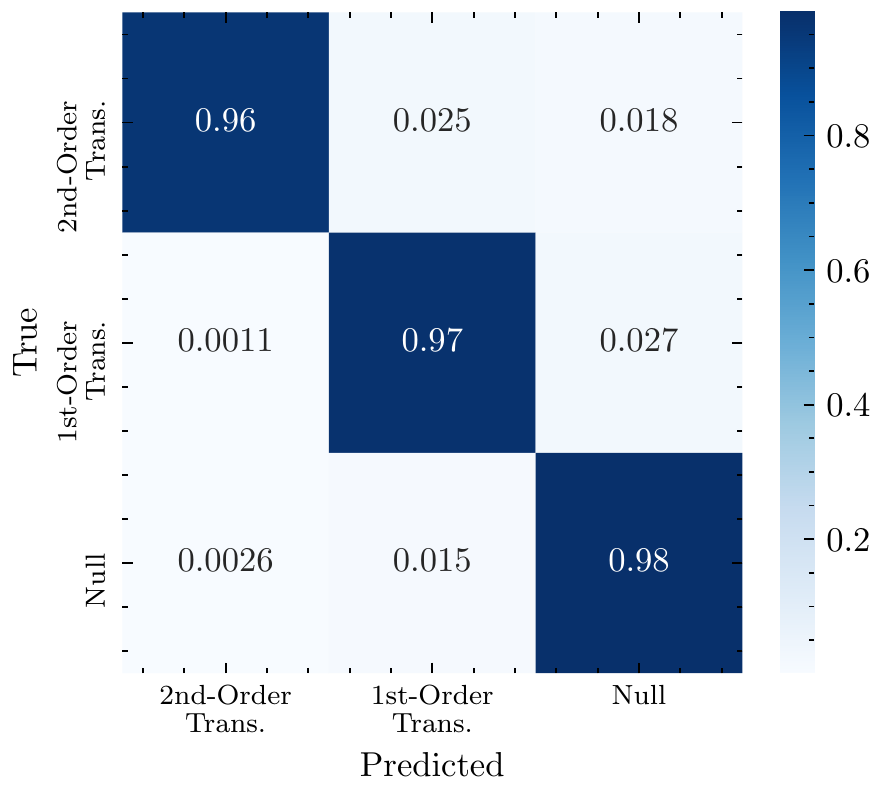}
\caption{Confusion matrix for results for a CNN-LSTM model trained to distinguish between first- and second-order Ising phase transitions, applied to a withheld test set of Ising data. Cell values represent normalized classification rates over the test set.}
\label{fig:Ising_ht_confusion_HP_10_all}
\end{figure}

The results presented in this paper strongly support the hypothesis of Hagstrom and Levin, even in the thermal systems on which we focus. The models trained on first-order Ising phase transitions perform comparably to corresponding models trained on second-order transitions when tested on both withheld Ising data (Fig. \ref{fig:var_ls_Ising_combine}) and other phase transition models (Fig. \ref{fig:Kefi_perc_ROC_combined_HP_1}). Moreover, while the theoretically-established critical precursors of spinodal instabilities are qualitatively indistinguishable from those of second-order phase transitions, classification results from our neural model show some promise of being able to distinguish between these phenomena. A separate model trained on a hybrid data set of first- and second-order Ising transitions was tasked with differentiating not only transition vs. null runs, but also with identifying the species of the upcoming transition. Results, plotted in Fig. \ref{fig:Ising_ht_confusion_HP_10_all}, show that the network was quite successful in learning this distinction. We note that the black-box nature of deep neural networks obfuscates the criteria by which the model makes this discrimination, so we cannot conclusively claim that it is making use of features of critical phenomena specifically (rather than other information content which delineates the classes), but nonetheless it offers an optimistic suggestion that such a distinction may be possible. 

\subsection*{Interpretations of differing model results}
Across the test cases presented, we sometimes observe significant discrepancies between the predictions of models trained on first- versus second-order phase transitions. In Fig. \ref{fig:CMIP_ROC_table}, for example, the second-order model achieves by far its highest classification accuracy on terrestrial (Lmon) variables, while the first-order model performs better on oceanic cryosphere (OImon) variables. Understanding these disparities is challenging, as the critical phenomena associated with both types of transition are theoretically predicted to be identical, and deep neural networks are famously opaque with respect to their decision criteria. We hypothesize that the primary driver of these differences lies in the properties of the phases to which the models are exposed during training: the second-order transitions occur between an ordered phase and a disordered one, whereas the first-order transitions occur between two ordered phases of opposite polarity. This likely results in models which are tuned to recognize transitions more closely resembling one or the other, which could account for the observed discrepancies (and probably also contributes to the results of the three-way classifier in Fig. \ref{fig:Ising_ht_confusion_HP_10_all}). Although we have largely confined our analysis to classifiers trained only on one order of transition at a time for the sake of improved interpretability, a combined training set may produce more robust predictions across a wider variety of test systems in future iterations of this model.

Some of the tests we present also exhibit disparities between classification accuracies of models trained on spatial and temporal statistical indicators. The likely mechanism underlying these differences is comparatively straightforward: even when a system is guaranteed to exhibit critical phenomena leading up to a phase transition, the magnitude and time/length scales of these behaviors depend on a variety of factors. Critical slowing down is observed through observation of a system recovering from small stochastic perturbations from equilibrium, but the properties of these perturbations and the specific morphology of the local potential well may vary from one system to another. Moreover, measurement of these phenomena must be carried out over some finite domain and resolution (in both spatial and temporal dimensions). This constrains observation sensitivity to a range of time and length scales, without any guarantee that these will align with the scales of the dynamics of interest. In the CMIP5 results presented in Fig. \ref{fig:CMIP_ROC_combined}, for example, the data analyzed is sampled annually at spatial intervals ranging from $1^\circ$ to $3^\circ$ latitude and longitude (depending on model). The critical transitions they contain, however, are presumed to arise across a variety of subsystems with dynamics playing out on very different scales. The comparatively strong performance of spatial indicators on abrupt shifts in the GFDL-ESDM2M data, or temporal indicators for the CSIRO-Mk3-6-0 data, can very likely be attributed to different predominant transition phenomena represented in these simulations, with critical phenomena which present more strongly in one domain or the other.

\subsection*{A note on selection bias}
Many approaches to early warning signal detection for critical transitions have been criticized for being calibrated on data exclusively from systems known to be prone to tipping events. This "prosecutor's fallacy," discussed in \cite{Boettiger2012}, leads to an inflated rate of false positive classifications when models are applied to generic out-of-sample data. The method we have presented in this paper shares this shortcoming---training sets are composed of transition and non-transition runs all derived from the 2D Ising model. Moreover, best practices in machine learning dictate that the populations of these classes should be balanced, which likely leads to an overrepresentation of critical transitions (assuming that non-criticality is the typical state in most systems of interest). As such, it should be expected that our models are somewhat overzealous in their positive classifications. Although the influence of the prosecutor's fallacy could be mitigated by populating the training set with any number of models not prone to phase transitions, we have chosen in this work to present models trained on highly minimal examples (i.e. only first- or second-order phase transitions in the 2D square Ising model). We take this approach to showcase the generalizability of the technique: even these simple, highly homogeneous data sets can be used to learn heuristics for universal critical phenomena which can be effectively transferred to much more complex systems. 

This strategy offers a useful and interpretable proof of concept, but it also violates the standard machine learning principle that models should be trained on the largest and most diverse data sets available. While we have prioritized interpretability in this work, we envision that a deployment-ready climate monitoring system based on this methodology would make use of a much more expansive collection of simulated systems, including a greater variety of both critical and non-critical dynamics. Further research will be needed to determine useful criteria for this diversity, but we hypothesize that such augmentations could yield considerable improvements of the model's generalizability and its susceptibility to false positives. For the time being, however, the outputs of these models should be interpreted as probabilities conditional on the assumption that the system in question is susceptible to critical transition behavior.

\section*{Outlook}

This paper is intended as a practical exhibition of deep learning methods for early detection of abrupt dynamical shifts due to phase transitions in spatiotemporal systems. This phase transition framework contrasts with previous deep learning approaches to EWS detection, which have focused on classifying local bifurcations based on scalar measurements on the dynamics of low-dimensional ODEs. Our results demonstrate the capacity of neural networks to learn generalizable features associated with critical phenomena of phase transitions. Moreover, the performance of models trained exclusively on spatial or temporal statistical indicators illustrates the potential of a higher-dimensional strategy: spatial features are shown to carry at least as much useful EWS information as temporal ones, information which is entirely lost when analyzing scalar time series.

This is an advancement of particular importance to climate applications, where variables of interest are often available via remote sensing data and recorded on a geospatial lattice. We envision the eventual use of models trained for the prediction of spatiotemporal tipping events applied in real time to global climate data across a spectrum of time and length scales, enabling an online warning system providing geolocated risk evaluations for critical phenomena in climate subsystems. Early warning of an oncoming tipping event is not, on its own, a particularly descriptive predictor of a system's future dynamics, but it can serve as a valuable supplement to existing analytical modes. Although the models we present cannot forecast dynamics past a tipping point, the knowledge of approaching criticality can enable modelers with expertise on the system in question to postulate a post-critical trajectory. Much work remains to further improve the robustness and generalizability of these models, but the ever-growing availability of data coupled with the capacity of neural networks for extraordinarily sophisticated pattern recognition suggests a promising future for this methodology.

\section*{Methods}

\subsection*{Training Data}
\label{sec:training_data}
To demonstrate the potential of generalizable early warning signal detection for spatiotemporal phase transitions, we train classifiers exclusively on simulations of the 2D square Ising model. This well-studied system undergoes a second-order phase transition between an ordered and disordered phase as its temperature is varied across a critical threshold, and a first-order phase transition from one ordered phase to another when an external magnetic field reverses polarity at fixed low temperature. Training data is generated by simulating $128 \times 128$ lattices with periodic boundary conditions using the Metropolis algorithm to stochastically update the state. Each Metropolis iteration flips a single, randomly selected lattice site with some probability. In order to obtain a time series for a comparatively smooth global evolution of the system, successive snapshots are sampled once every $10^4$ Metropolis iterations. Two classes of runs are performed: transition runs, in which temperature $T$ (or external field $h$) is varied linearly through its critical regime, and null runs, in which $T$ ($h$) is varied linearly but held far from criticality. In all cases, the system is initialized far from its critical regime (relative to the magnitude of steady-state thermal fluctuations).

For the sake of generality, steps are taken to break the natural symmetries of the system: coupling strengths $J_{ij}$ between each pair of neighboring lattice sites are randomly sampled from a normal distribution, and the structure of the lattice itself is modified by deleting sites from randomly selected elliptical subregions. This serves to improve generalizability of the trained models by producing more diverse critical phenomena (albeit within a narrow and fairly homogeneous subset of all possible phase transitions). A further obstacle to generalization is posed by the binary nature of the Ising model, in which lattice sites are constrained to values of $\pm 1$. This is accounted for by coarse grain averaging in space and time to achieve a smoother distribution of states (with the specific parameters of the coarse graining randomized to promote more diverse representation of time and length scales).

To preprocess data for training, Ising runs are first truncated to end 100 time steps before the observed transition. For second-order Ising transitions, we use the analytical result of $T_c = 2J/\log(1+\sqrt{2})$. This result is not exact for these simulations due to the randomized coupling strengths and lattice site deletions, but these perturbations are sufficiently minor that it provides a workable approximation (using the mean coupling strength for $J$). For first-order transitions there is no precise critical value for $h$ at which phase overturning takes place (due to the temporary metastability of the starting phase), so the transition time is defined by numerically computing the inflection point of the net magnetization of the lattice. Null runs, in which no transition takes place, are truncated to match their length distribution to that of the critical runs. Each run is derived from a $9\times 9$ spatial grid measured over up to 600 time steps, which is obtained by spatially coarse-graining the Ising lattice by a randomized factor between 4 and 15 (sampled from a correspondingly larger subregion of the lattice). This serves to smooth out the discrete binary states of individual lattice sites. Run durations are also randomized (between 100 and 600 time steps), and padded from the left by zeros to maintain the fixed input dimension of the neural network.

In the interest of avoiding over-parameterization in the neural model, an additional dimensionality-reduction step is applied to this data: a time series in 81 dimensions is reduced to 12 by computing aggregate statistics which are theoretically predicted to carry information relevant to critical phenomena. Specifically, variance, skewness, kurtosis, and 1-, 2-, and 3-autocorrelation are computed in both the space and time domains. The spatial quantities are computed independently for each observed time step, while the temporal quantities are computed across a sliding window with randomized width (between $10\%$ and $40\%$ of the total domain) and averaged over the spatial dimensions. Additionally, following the conventional wisdom of traditional methods for EWS detection, temporal statistics are computed on time series residuals obtained by subtracting off a Gaussian filter moving average. Importantly, this filter must be applied after the truncation of the time domain, so as not to allow information from after the critical point to indirectly influence earlier time points via the Gaussian kernel. Finally, time series are normalized to have mean 0 and unit variance, as is standard for inputs to neural models to be trained by gradient descent methods. 

We note that this is not the only means by which dimensionality reduction might be accomplished; the neural network architecture could, for example, be modified to implement a learned ``bottleneck" module to step down the input dimension while retaining useful information. Indeed, neural autoencoders are often used to great effect to obtain a compressed representation of data for classification tasks. For the sake of simplicity we have elected not to take this approach in this paper, favoring more interpretable pre-defined statistical indicators (which carry the added benefit of separating spatial and temporal features), but in future work we plan to explore alternative techniques.

\subsection*{Model Design}
\label{sec:model_design}
\begin{figure}[!ht]
\centering
\includegraphics[width=0.8\columnwidth]{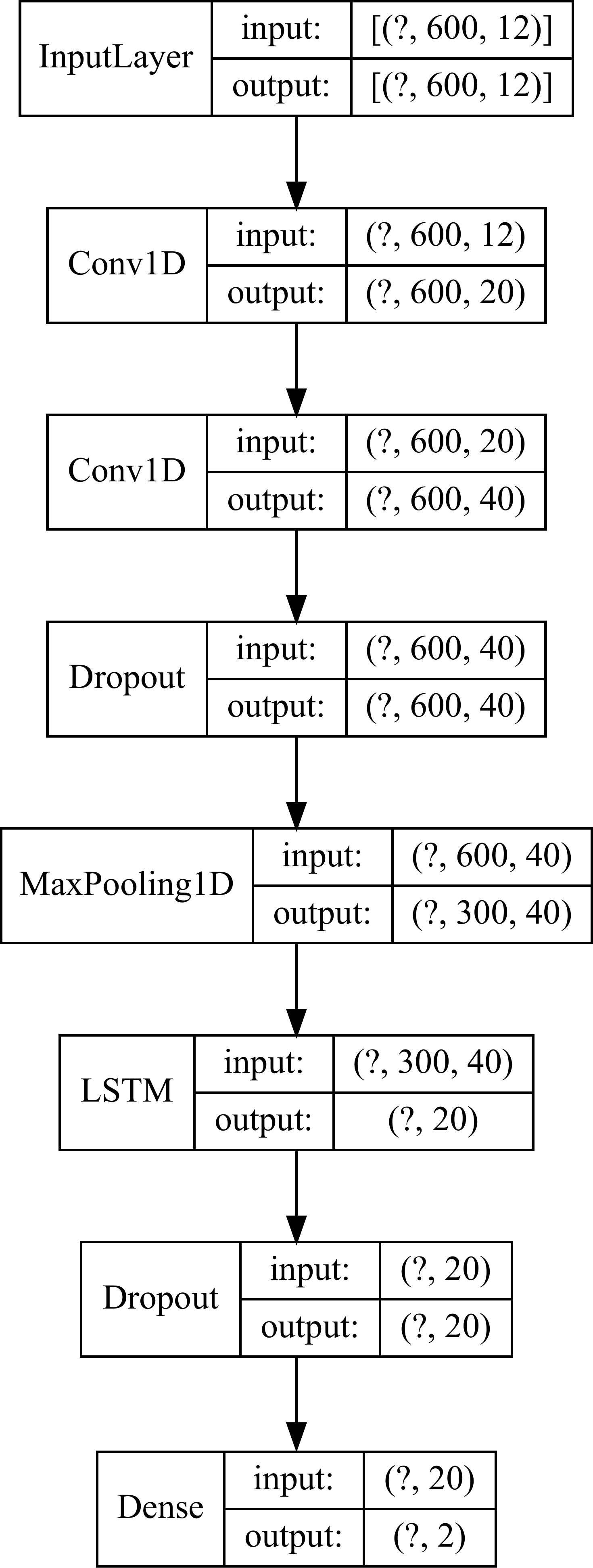}
\caption{This schematic depicts the architecture of the neural network. The leading ``?" index of each layer shape refers to batch size, which is a user-defined parameter.}
\label{fig:model_schematic}
\end{figure}

The deep neural network for early warning signal detection is trained for the simple binary classification task of differentiating between Ising time series data with and without a phase transition imminently approaching. Taking inspiration from Bury et. al. \cite{Bury2021}, we employ a CNN-LSTM architecture in which a convolutional component feeds into a long short-term memory (LSTM) module. Convolutional methods represent the state of the art in neural time series classification \cite{Fawaz2019}, and the inclusion of an LSTM suits the problem of EWS detection: theory of critical phenomena predicts early-warning indicators which manifest as deviations from some baseline behavior of the system, and recurrent neural networks contain hidden state variables which can act as a ``memory bank" to encode a representation of this baseline. Specifically, the model contains two convolutional layers, each with 20 one-dimensional filters of width 8. This is followed by a 10\% dropout layer and a max pooling layer with stride 2. The output of this feeds into a single LSTM module containing 20 memory cells followed by a second dropout layer and finally a dense softmax layer into two output dimensions for the binary classification. A schematic of this architecture is presented in Fig. \ref{fig:model_schematic}. This architecture was validated by comparison to analogous models containing only CNN or only LSTM components, which achieved worse classification accuracy, and models with additional CNN or LSTM modules, which exhibited signs of overfitting to the test data. Other hyperparameters were subjected to limited tuning (e.g. the number of neurons in each CNN/LSTM module was also selected to maximize validation accuracy without overfitting), but the extreme success of the model in classifying withheld test data obviated the need for extensive refinement. Separate models were trained for inputs containing the 6 spatial indicators, 6 temporal indicators, or a combination of the two (input dimension 12), with all other architectural properties held constant. 

\backmatter

\section*{Declarations}

\subsection*{Funding}
The research was supported by NSERC Discovery Grants to MA (5006032-2016) and CTB (5013291-2019), and a DARPA Artificial Intelligence Exploration Opportunity Grant to MA, CTB and TL (PA-21-04-02-ACTM-FP-012).

\subsection*{Acknowledgments}

We thank Rick Hennekam and Gert-Jan Reichart for providing the anoxia dataset. We acknowledge the World Climate Research Programme’s Working Group on Coupled Modelling, which administrates the CMIP5 database, as well as the contributors of the GFDL-ESM2M, FGOALS-g2, CSIRO-Mk3-6-0, and CMCC-CESM model data.

\subsection*{Data and code availability}
Python code to generate Ising model training data and synthetic test simulations is available on the author's Github page: \url{https://github.com/dylewsky/phase_transition_EWS}.

Climate simulations from the CMIP5 collaboration are made available by the World Climate Research Programme and the Program for Climate Model Diagnosis \& Intercomparison. Details can be found at \url{https://pcmdi.llnl.gov/mips/cmip5/}.

\onecolumn

\bibliography{ews_paper_v4}

\end{document}